\documentclass[aps,prl,twocolumn,superscriptaddress,longbibliography]{revtex4-1} 
\pdfoutput=1
\usepackage[version=3]{mhchem} % Formula subscripts using \ce{}
\usepackage[dvips]{epsfig}
\usepackage{graphicx}  % needed for figures
\usepackage{dcolumn}   % needed for some tables
\usepackage{bm}        % for math
\usepackage{amssymb}   % for math
\usepackage{amsmath} % for math
\usepackage{bm} % For bold math
\usepackage{color}
\usepackage[dvipsnames]{xcolor}

\begin{document}

\author{Marcelo Wu}
\thanks{These authors contributed equally to this work.}
\affiliation{Department of Physics and Astronomy \& Institute for Quantum Science and Technology, University of Calgary, Calgary, AB, T2N 1N4, Canada}
\affiliation{National Institute for Nanotechnology, Edmonton, AB, T6G 2M9, Canada}

\author{Nathanael L.-Y. Wu}
\thanks{These authors contributed equally to this work.}
\affiliation{Department of Physics and Astronomy \& Institute for Quantum Science and Technology, University of Calgary, Calgary, AB, T2N 1N4, Canada}
\affiliation{National Institute for Nanotechnology, Edmonton, AB, T6G 2M9, Canada}

\author{Tayyaba Firdous}
\thanks{These authors contributed equally to this work.}
\affiliation{National Institute for Nanotechnology, Edmonton, AB, T6G 2M9, Canada}
\affiliation{Department of Physics, University of Alberta, Edmonton, AB, T6G 2E9, Canada}

\author{Fatemeh Fani Sani}
\affiliation{National Institute for Nanotechnology, Edmonton, AB, T6G 2M9, Canada}
\affiliation{Department of Physics, University of Alberta, Edmonton, AB, T6G 2E9, Canada}

\author{Joseph E. Losby}
\affiliation{National Institute for Nanotechnology, Edmonton, AB, T6G 2M9, Canada}
\affiliation{Department of Physics, University of Alberta, Edmonton, AB, T6G 2E9, Canada}

\author{Mark R. Freeman}
\email{freemanm@ualberta.ca}
\affiliation{National Institute for Nanotechnology, Edmonton, AB, T6G 2M9, Canada}
\affiliation{Department of Physics, University of Alberta, Edmonton, AB, T6G 2E9, Canada}

\author{Paul E. Barclay}
\email{pbarclay@ucalgary.ca}
\affiliation{Department of Physics and Astronomy \& Institute for Quantum Science and Technology, University of Calgary, Calgary, AB, T2N 1N4, Canada}
\affiliation{National Institute for Nanotechnology, Edmonton, AB, T6G 2M9, Canada}

\title{Nanocavity optomechanical torque magnetometry and radiofrequency susceptometry}

\maketitle

%%% INTRO %%% 
\noindent
\textbf{Nanophotonic optomechanical devices allow observation of nanoscale vibrations with sensitivity that has dramatically advanced metrology of nanomechanical structures \cite{ref:li2008hof, ref:liu2012wcs, ref:anetsberger2010mnm, ref:krause2012ahm, ref:kim2013nto, ref:wu2014ddo, ref:li2014ops, ref:chan2011lcn, ref:zhang2016cm} and has the potential to impact studies of nanoscale physical systems in a similar manner \cite{ref:rugar2004ssd, ref:bleszynskijayich2009pcn}. Here we demonstrate this potential with a nanophotonic optomechanical torque magnetometer and radiofrequency (RF) magnetic susceptometer. Exquisite readout sensitivity provided by a nanocavity integrated within a torsional nanomechanical resonator enables observations of the unique net magnetization and RF-driven responses of single mesoscopic magnetic structures in ambient conditions. The magnetic moment resolution is sufficient for observation of Barkhausen steps in the magnetic hysteresis of a lithographically patterned permalloy island \cite{ref:burgess2013qmd}. In addition, significantly enhanced RF susceptibility is found over narrow field ranges and attributed to thermally assisted driven hopping of a magnetic vortex core between neighboring pinning sites \cite{ref:compton2006dpm}. The on-chip magneto-susceptometer scheme offers a promising path to powerful integrated cavity optomechanical devices for quantitative characterization of magnetic micro- and nanosystems in science and technology.}

Torque magnetometry has seen recent resurgence owing to miniaturization of mechanical devices \cite{ref:moreland2003mif}. The high detection sensitivity of resonant nanomechanical torque sensors has allowed for minimally-invasive observations of magnetostatic interactions and hysteresis in a variety of magnetic materials including thin films \cite{ref:lim2014mis}, mesoscale confined geometries that are deposited \cite{ref:davis2010nrt} or epitaxially grown \cite{ref:losby2014nas}, and small aggregates of nanoparticles \cite{ref:firdous2015ntm}.  Going beyond the static limit, nanomechanical torque magnetometry has been extended to timescales allowing for detection of slow thermally-activated dynamics \cite{ref:burgess2013qmd}, AC susceptibility \cite{ref:losby2014nas}, and magnetic resonance \cite{ref:ascoli1996mdm, ref:losby2015tmr}.   

%%%%% FIGURE 1 %%%%%
\begin{figure}[tb]
\begin{center}
\epsfig{figure=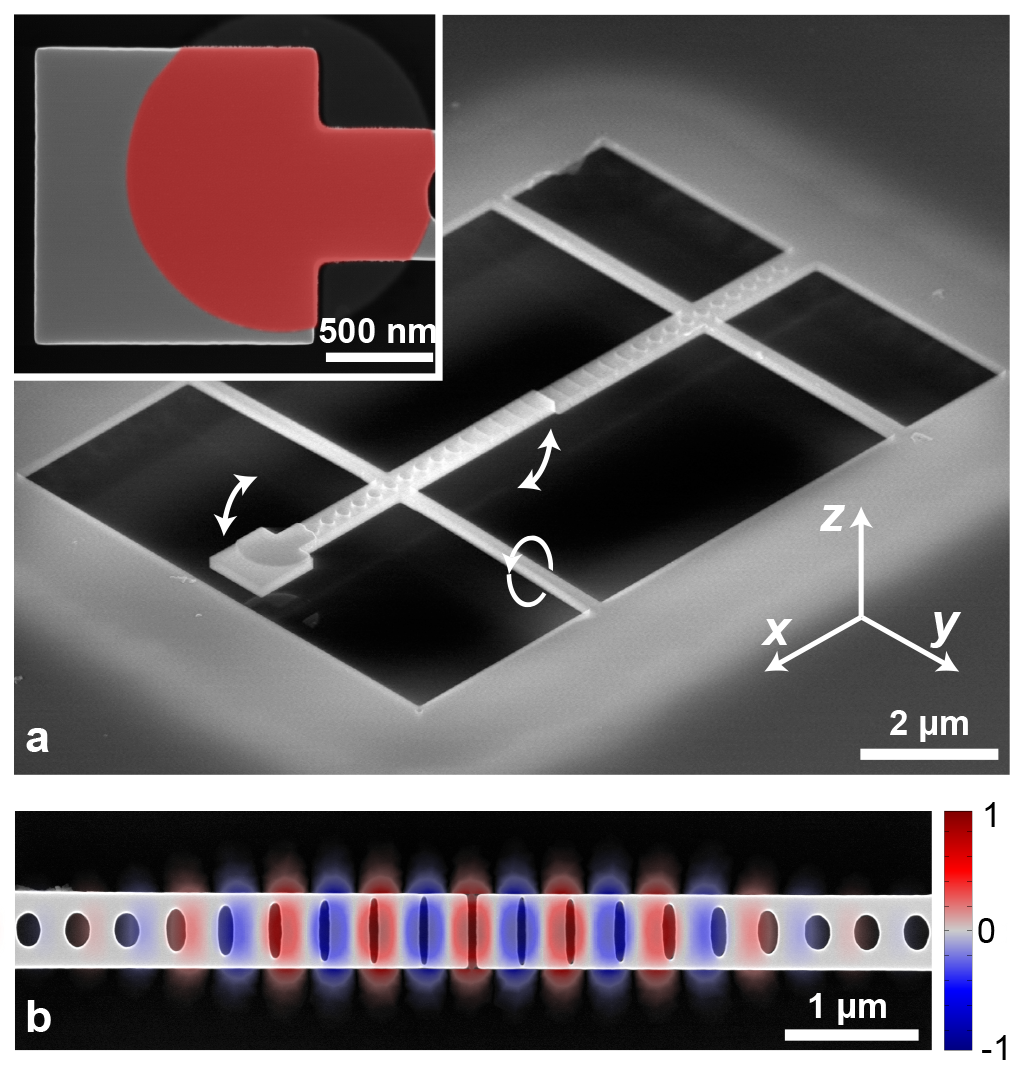, width=1\linewidth}
 \caption{\textbf{Split-beam nanocavity.} \textbf{a}, Tilted scanning electron micrograph (SEM) of a split-beam cavity optomechanical torque sensor supporting a 40 nm thick permalloy island (highlighted in red in the inset). \textbf{b}, Top-view SEM of the nanocavity overlaid with a finite element simulation (COMSOL) of the normalized field distribution $E_\text{y}$ of its optical mode.}
\label{fig:Device}
\end{center}
\end{figure}

%%%%% FIGURE 2 %%%%%
\begin{figure*}[htb]
\begin{center}
\epsfig{figure=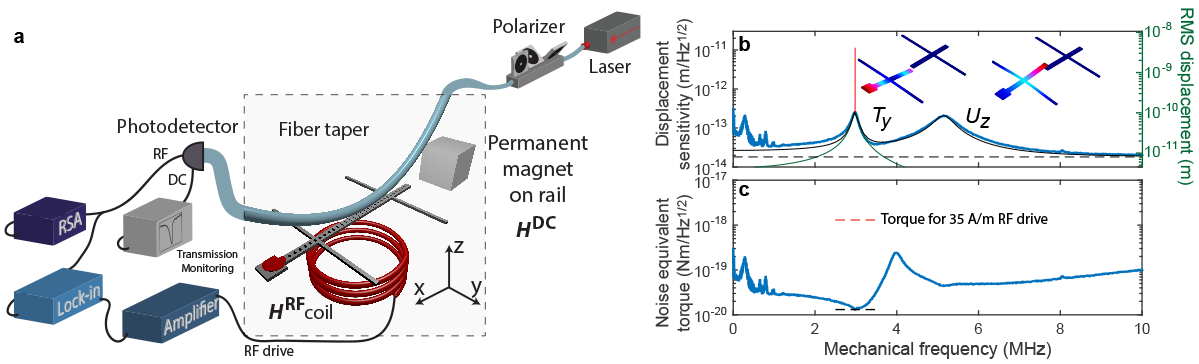, width=1\linewidth}
 \caption{\textbf{Measurement setup and spectral response.} \textbf{a}, Experimental setup for nanocavity optomechanical torque magnetometry measurements (not to scale). All measurements are performed in an ambient nitrogen purged environment (gray region). A dimpled fiber taper is used to probe the optomechanical nanocavity. A permanent magnet with adjustable position provides varying static magnetic fields. The lock-in amplifier reference is power amplified and sent to coils below the device to creates an RF magnetic field in the $\hat{z}$-direction. \textbf{b}, Displacement density (left axis) from the real-time spectrum analyzer (RSA) showing thermally driven mechanical modes $T_\text{y}$  and $U_\text{z}$ (blue) and the magnetically driven signal (narrow peak highlighted in red) generated by a magnetic driving field $H^\text{RF}_\text{z}$ of 35 A/m applied with the permalloy island magnetization saturated by $H^\text{DC}_\text{x}$. Black lines are fits to the $T_\text{y}$ and $U_\text{z}$ Lorentzian-shaped peaks (solid line) and the measurement noise floor (dashed line). The green curve (right axis) indicates  the predicted RMS displacement of the $T_\text{y}$ resonance in the presence of 35 A/m RF magnetic field as a function of frequency. Insets show simulated displacement profiles of $T_\text{y}$  and $U_\text{z}$.  \textbf{c}, Torque equivalent noise of the thermomechanical displacement signal in \textbf{b}. The red dotted line indicates the predicted torque in the presence of a 35 A/m $H_\text{z}^\text{RF}$ field, and is labeled by the values on the left axis assuming a 1 s integration time.}
\label{fig:Sensitivity}
\end{center}
\end{figure*}

This powerful technique relies upon detection of the deflection of a mechanical element by angular momentum transfer originating from magnetic torques $\boldsymbol{\tau} = \mu_\text{0} \bm{m} \times \bm{H}$, generated as the magnetic moments in the system, $\bm{m}$, experience an orthogonally-directed component of the applied magnetic field, $\bm{H}$. So far, improvements to torque magnetometers have been driven primarily by enhancements to the response of nanomechanical resonators resulting from their low mass and high mechanical quality factor ($Q_\text{m}$).  Readout of magnetically driven motion has involved detection through free-space optical interferometric methods with very low optical quality factor ($Q_\text{o} \approx 1$) Fabry-Perot cavities formed between the nanomechanical resonator its supporting substrate \cite{ref:davis2010nrt}. However, as device dimensions scale down, and the number of magnetic spins become too small or the dynamics too fast, the mechanical deflections become more difficult to detect. Migration to a more sensitive readout scheme is essential. The integration of a nanoscale optical cavity offers a natural path for improvement.

Nanocavity-optomechanical devices enhance mechanical detection sensitivity by confining light to high-$Q_\text{o}$ modes localized within the nanomechanical resonator.  They have been exploited for metrology applications including force and displacement detection \cite{ref:liu2012wcs, ref:anetsberger2010mnm}, inertial sensing \cite{ref:krause2012ahm}, torque sensing \cite{ref:kim2013nto, ref:wu2014ddo, ref:li2014ops}, and  observation of mechanical quantum fluctuations \cite{ref:chan2011lcn}. Recently, microscale ($\sim$ 10 -- 100 $\mu$m) cavity optomechanical devices were combined with magnetostrictive materials to create external magnetic field sensors \cite{ref:forstner2014uom}. Nanophotonic optomechanical devices with sub-wavelength dimensions have tremendous potential to impact mechanical sensing of the microscopic electronic and magnetic dynamics of meso- and nanoscale systems typically readout using conventional optical methods \cite{ref:rugar2004ssd, ref:bleszynskijayich2009pcn}.  In this Letter, we apply a nanocavity optomechanical sensor to a nanoscale condensed matter system for the first time and demonstrate that torque magnetometry can be performed with sufficient sensitivity for detection of Barkhausen features that were previously undetected in ambient conditions. We use this device to demonstrate a new form of nanomechanical RF susceptometry,  and observe  enhanced magnetic susceptibility associated with single pinning and depinning events that can increase the torque magnetometer sensitivity by over an order of magnitude.

%%% DESCRIPTION OF DEVICE %%%  

The high sensitivity of nanocavity optomechanical devices arises from a combination of large optomechanical coupling, large mechanical resonator susceptibility (low mass $m_\text{eff}$, and large $Q_\text{m}$) and sharp optical cavity response (large $Q_\text{o}$). The device employed here, an example of which is shown in Fig.\ \ref{fig:Device}a, is designed with these properties in mind. Referred to as a split-beam nanocavity (SBC) \cite{ref:wu2014ddo}, it consists of two suspended silicon photonic crystal nanobeams -- one anchored in three sections, and the other ``moving nanobeam'' anchored by two supports. It supports an optical mode whose field, shown in Fig.\ \ref{fig:Device}b, is confined to the central gap region and has a high-$Q_\text{o}$ ($\sim 5,000$ for the device studied here) owing to careful mode matching between the local field supported by the gap with the field in the elliptical holes of the nanobeams \cite{ref:hryciw2013ods}. Vibrations of nanobeam mechanical resonances modulate both the gap width and the distance between the SBC and the fiber taper waveguide used to evanescently couple light into and out of the nanocavity, as illustrated in the experimental setup in Fig.\ \ref{fig:Sensitivity}a (details in Methods).  Of particular interest for torque magnetometry is the torsional resonance $T_\text{y}$ of the moving nanobeam, where the nanobeam ends move anti-symmetrically out-of-plane as shown schematically in Fig.\ \ref{fig:Device}a. This low mass ($m_\text{eff}\sim 1\,\text{pg}$) resonance (frequency $\omega_\text{m}/2\pi \sim 3$ MHz) can be efficiently excited by nanoscale sources of torque coupled to the SBC.

The interaction between nanobeam motion and nanocavity optical dynamics is characterized by the optomechanical coupling coefficient $g_\text{om}$ (see \href{\SMlink}{Supplement} \S1). Detection of the vertical motion of the $T_\text{y}$ resonance relies on the dispersive optomechanical interaction between the SBC and the fiber taper. The fiber taper renormalizes the nominally symmetric nanocavity field and induces $g_\text{om}/2\pi$ in the gigahertz per nanometre range \cite{ref:hryciw2015tno}. The displacement sensitivity of the fiber coupled SBC device can be calibrated by measuring the optomechanically transduced thermal motion, as shown in Figs.\ \ref{fig:Sensitivity}b,c. For typical operating conditions, it is in the tens of fm$/\sqrt{\text{Hz}}$ range with an equivalent torque of $1.3 \times 10^{-20} \text{N m}/\sqrt{\text{Hz}}$ (details in \href{\SMlink}{Supplement} \S2 and \S3). All measurements are performed in ambient conditions, resulting in  $Q_\text{m} < 100$ because of viscous air damping.

Nanocavity torque magnetometry can be performed by actuating the $T_\text{y}$ mode with a magnetic field $\bm{H}$ that interacts with a magnetic moment $\bm{m}$ on the nanobeam \cite{ref:rigue2012tmt, ref:losby2010ntm}. Here we investigate the magnetic properties of a ferromagnetic thin-film permalloy island integrated onto a rectangular pad at the end of the moving nanobeam, shown in the inset in Fig.\ \ref{fig:Device}a (see Methods). When an in-plane static field $H^\text{DC}_\text{x}$ is applied, the permalloy becomes magnetized with net moment $m_\text{x} (H_\text{x}^\text{DC})$ along the field $\hat{x}$-direction. By applying an additional RF field $H^\text{RF}_\text{z}$ directed in the out-of-plane $\hat{z}$-direction, a magnetic torque $\tau_\text{y}$ is generated proportional to $m_\text{x}$ and directed along the torsion rod supporting the moving nanobeam.  When the RF field is applied at the $T_\text{y}$ resonance angular frequency $\omega_\text{m}$, the resulting driven beam displacement can be detected optomechanically from the nanocavity optical response. A typical signal is shown in the spectral domain in Fig.\  \ref{fig:Sensitivity}b, where it clearly emerges as a sharp peak at $\omega_\text{m}$ far above the thermomechanical noise. {Note that while this device has the largest optomechanical magnetic transduction of those fabricated for this study, other devices (see \href{\SMlink}{Supplement} \S7) were observed to display behaviour similar to that described throughout this letter.}  

%%%%% FIGURE 3 %%%%%
\begin{figure}[htb]
\epsfig{figure=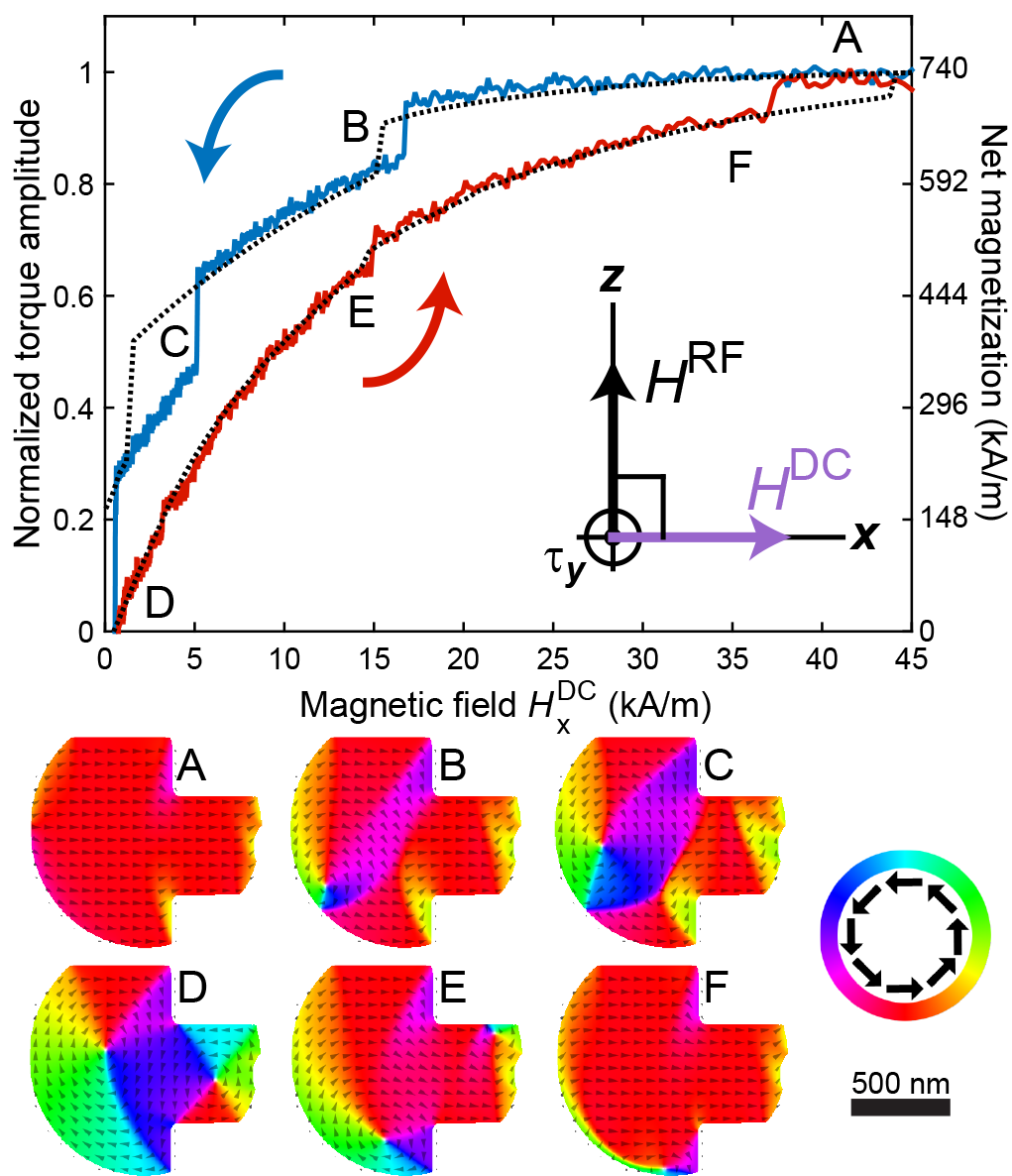, width=0.5\textwidth}
 \caption{\textbf{Magnetic hysteresis of the permalloy island.} Magnetization response of the permalloy element with varying applied DC field along $\hat{x}$ (5 runs averaged). The RF drive field is $H^\text{RF}_\text{z} = 35$ A/m.  The solid blue trace is a decreasing field sweep and the solid red trace is an increasing field sweep. Results from micromagnetic simulations of the permalloy island (highlighted in red in the SEM inset in Fig. \ref{fig:Device}a and also used as the simulation mask) are plotted with black dashed lines. Bottom: simulated magnetization textures at different points in the hysteresis loop. The color wheel shows the in-plane direction of magnetization, with red parallel to the applied DC field. }
\label{fig:Hysteresis}
\end{figure}

%%%%% FIGURE 4 %%%%%
\begin{figure*}[htb]
\epsfig{figure=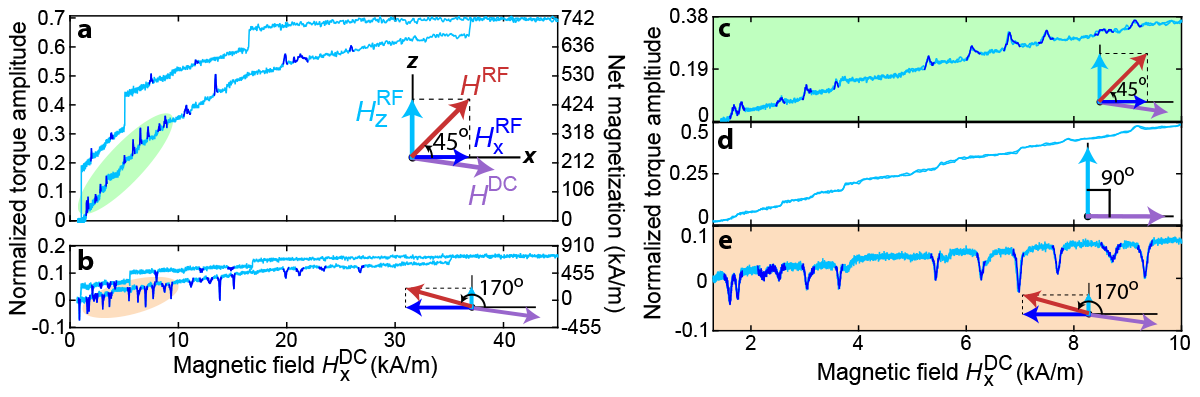, width=1.0\linewidth}
\caption{\textbf{Enhanced room temperature magnetic susceptibility at Barkhausen steps}. Hysteresis sweep with $H^\text{RF}$ set to \textbf{a}, 45$^o$ (equal and same sign $x$ and $z$ components) and \textbf{b}, 170$^o$ (opposite signed $x$ and $z$ components). A select number of upward and downward peaks have been highlighted in blue to show contribution to torque from susceptibility. Low field $H^\text{DC}_\text{x}$ single forward and backward sweeps at three $H^\text{RF}$ positions: \textbf{c} 45$^\text{o}$, \textbf{d} 90$^\text{o}$, and \textbf{e} 170$^\text{o}$. In all cases, the drive field $H^\text{RF} = 35$ A/m.}
\label{fig:OffsetPeaks}
\end{figure*}

%%% HYSTERESIS MEASUREMENTS %%%
To perform magnetometry on the permalloy island, hysteresis loops were measured by varying $H^\text{DC}_\text{x}$ via translation of the permanent magnet while recording the optomechanically transduced RF signal for fixed $H^\text{RF}_\text{z}$ using the lock-in amplifier. Figure \ref{fig:Hysteresis} shows the torque signal normalized to the value at saturation, with the corresponding scale for the net magnetization on the right axis. Beginning at high field (blue curve in Fig.\ \ref{fig:Hysteresis}), the magnetization was nearly saturated (section of the curve labeled A; the correspondingly lettered frames in the bottom section of Fig.\ \ref{fig:Hysteresis} are representations of the spin textures from micromagnetic simulation, see Methods). As the field decreases, three large discontinuities in the  net moment inferred from the optomechanical signal are observed and correspond to irreversible changes in the spin texture, beginning with nucleation of a magnetic vortex with an out-of-plane core surrounded by in-plane curling magnetization (section B of the curve).  As the DC field is further decreased, the vortex core translates towards the center of the element until an intermediate texture arises, featuring pronounced closure domains along the short edges perpendicular to the applied field (section C).  The transition near zero field forms a two-vortex state, shown in frame D of Fig.\ \ref{fig:Hysteresis}, where the permalloy island's mushroom-like shape supports a Landau state in the stem (right side) and a distorted circular vortex in the cap (left side) \cite{ref:cowburn1999scn} in keeping with the demagnetizing energetic preference for the moments near edges to be nearly tangential to the boundaries.  When $H^\text{DC}_\text{x}$ is subsequently increased (red curve), the net moment increases monotonically with applied field.  In section D of the simulation the two vortex cores move in opposite directions perpendicular to the field as the two circulations have opposite chiralities in this instance.  The simulation frames E and F show the spin configurations just before each individual vortex core annihilates, after being pushed too close to the edge to remain stable by the field-increasing sweep.  The simulated hysteresis loop (black dashed line) shows good qualitative agreement with observation, with the difference in the transition field values in part due to the simulations having been performed without including thermal energy.

%%% BARKHAUSEN JUMPS %%%
Figure \ref{fig:OffsetPeaks} demonstrates the ability of the nanocavity optomechanical torque sensor to capture, in high resolution measurements, fine structure in the hysteresis that is the fingerprint of intrinsic disorder unique to a given permalloy island, and can not be predicted by the idealized micromagnetic simulations described above.  The high energy density of vortex cores make them susceptible to pinning at imperfections (surface roughness and grain boundaries) in the polycrystalline island. With diameters on the order of ten nanometers, the cores finely probe the magnetic landscape as their positions change with applied field \cite{ref:burgess2013qmd}.  Pinning and depinning events are captured as Barkhausen steps, with notable reductions in slope of the hysteresis curve seen whilst cores are pinned.  Figure \ref{fig:OffsetPeaks} shows a rich spectrum of repeatable events whose character varies depending on the orientation of $\bm{H}^\text{RF}$, as indicated in each of Figs.\ \ref{fig:OffsetPeaks}a--e. Repeatable events for $\bm{H}^\text{RF}$ perpendicular to the permalloy film (i.e., along $\hat{z}$) visible in section D of Fig.\ \ref{fig:Hysteresis} are shown in close-up in Fig.\ \ref{fig:OffsetPeaks}d. If the applied field is kept below the first vortex core annihilation field, curves like Fig.\ \ref{fig:OffsetPeaks}d show distinct steps without hysteresis when the field strength is ramped down.  The absence of any minor hysteresis at each step is the result of very rapid (in comparison to the measurement bandwidth) thermally-activated hopping between neighboring pinning centers \cite{ref:burgess2013qmd, ref:compton2006dpm}, such that the apparatus records a temporal average weighted by the relative dwell times in the two sites.  

For non-normal $\bm{H}^\text{RF}$, the nanocavity torque sensor can function as a susceptometer that probes RF magnetic susceptibility and provides new insight into the properties of the pinning processes. For these measurements, an in-plane $\hat{x}$-component of the RF field (parallel to the nominal DC field direction) is introduced by tuning the relative RF coil position off-centre to the device (details in \href{\SMlink}{Supplement} \S5). Note that adjusting the relative chip-coil position is simplified experimentally by the ambient operating conditions and fiber-based readout. A small out-of-plane DC field $H_\text{z}^\text{DC}$ combines with the oscillating field $H_\text{x}^\text{RF}$ to generate torque in the $\hat{y}$-direction proportional to the in-plane susceptibility.  Signals recorded using both $\hat{z}$ and $\hat{x}$ components of RF drive contain both torque contributions: from the net moment along $\hat{x}$ ($\propto m_\text{x}^\text{DC}H_\text{z}^\text{RF}$) and from the RF susceptibility along $\hat{x}$ ($\propto \chi^\text{RF}_\text{x}H_\text{x}^{RF}H_\text{z}^\text{DC}$), where $\bm{\chi}$ is the magnetic susceptibility tensor of the permalloy island (see \href{\SMlink}{Supplement} \S5).  

Figure \ref{fig:OffsetPeaks}a,b shows the full hysteresis loops for two different RF field orientations: 45$^\text{o}$ (\ref{fig:OffsetPeaks}a) and 170$^\text{o}$ (\ref{fig:OffsetPeaks}b) anticlockwise from horizontal. The torque values remain normalized to the 90$^\text{o}$ orientation. Corresponding close-ups of the low-field sections are shown in Figs.\ \ref{fig:OffsetPeaks}c,e.  The peaks and dips newly found in the data are RF susceptibility signatures arising when the energy barrier between neighboring pinning sites is small enough that the in-plane RF field is able to drive the core synchronously back-and-forth. To the best of our knowledge, these measurements are the first report of RF susceptibility due to the Barkhausen effect at the single pinning event level,  though averaged events have been studied previously \cite{ref:abulibdeh2010ddg}. Note that the larger transitions between spin textures in the main loop are irreversible and therefore exhibit no accompanying RF susceptibility features.  The effective susceptibility (calculated in \href{\SMlink}{Supplement} \S6) $\delta\bm{m}/\delta\bm{H}$ will be largest when the RF drive amplitude is just above the threshold required for a synchronous response, where the ratio of $\delta\bm{m}$ (set to first approximation by the moment change at the Barkhausen jump) to $\delta\bm{H}$ is largest.  Observed enhancements of up to $25$ times over the susceptibilities while the core is pinned suggest applications to RF susceptibility engineering in applications such as field-sensing magnetometry and detecting small volumes of magnetic material. 

Both the ratios of amplitudes and the relative signs of the net moment and susceptibility contributions in Fig.\ \ref{fig:OffsetPeaks} are consistent with the changes of the RF field direction.  Implementation of a scheme with independent control of RF field components will enable quantitative separation of the susceptibility and magnetometry components through $\pi$ phase shifts of individual RF drives without changing anything else, providing further confirmation of the phenomena reported above. A proof-of-principle demonstration of our ability to probe different components of the susceptibility through reconfiguration of the RF field direction is presented in \href{\SMlink}{Supplement} \S7, where the off-diagonal susceptibility of the pinning events is detected in this way. Given the already important role of thermally-driven rapid hopping in eliminating observed minor hysteresis at Barkhausen steps \cite{ref:burgess2013qmd}, the synchronization must be thermally-assisted. Operating the device at low temperature in future work is required to search for threshold behavior.

%%% CONCLUSION %%%

The ability of the optomechanical nanocavity to detect nanoscale magnetic phenomena arises from its torque sensitivity of $1.3 \times 10^{-20} \text{N} \, \text{m}/\sqrt{\text{Hz}}$, which at field strengths on the order of Earth's field (44 -- 60 $\mu$T), corresponds to magnetic moment sensitivity of $(2.4 \pm 0.4)\times 10^7 \, \mu_\text{B}$. A minimum detectable volume of magnetic material of $0.015 \pm 0.005 \, \mu \text{m}^3$ is calculated for the largest susceptibility enhancement; increasing the RF drive would allow for measurement of even smaller volume samples. Compared to previous nanoscale torque magnetometry devices \cite{ref:davis2010nrt, ref:burgess2013qmd, ref:losby2015tmr} reliant on free-space reflectometry and vacuum or cryogenic operation, this device is of comparable or better sensitivity despite operating in ambient conditions. Furthermore, its relatively low $Q_\text{m}$ and MHz operating frequency in-principle allows MHz bandwidth excitation and detection. Among nanoscale optomechanical torque metrology devices, the demonstrated sensitivity is only surpassed by systems operating in vacuum \cite{ref:li2014ops} or cryogenic conditions \cite{ref:kim2016asq}, none of which have yet been used for magnetometry or to probe nanoscale condensed matter systems. 

Notwithstanding the practical advantages enabled by ambient conditions operation, vacuum and low temperature $T_\text{b}$  will reduce the thermal force fluctuations that scale with $\sqrt{T_\text{b}/Q_\text{m}}$ and limit sensitivity \cite{ref:wu2014ddo}. For example, $Q_\text{m} \sim 10^3 - 10^4$ for similar SBC devices has been observed in vacuum 
 \cite{ref:wu2014ddo}, and $Q_\text{m} \sim 10^5$ for silicon zipper nanocavity devices has been observed at liquid helium temperatures \cite{ref:safavinaeini2013sls}. This indicates that a $10^4$ improvement in the thermally limited sensitivity may be within reach. Even a modest improvement in sensitivity by an order of magnitude, in combination with a maximum driving field of 1 kA/m, could produce magnetic moment sensitivities below $2 \times 10^{5}\mu_{B}$ \cite{ref:kim2013nto}, enabling nanomagnetism-lab-on-chip studies of a wide range of systems \cite{ref:rugar2004ssd, ref:bleszynskijayich2009pcn}.

In conclusion, we have experimentally demonstrated nanocavity optomechanical detection for torque magnetometry and RF susceptometry. The device presented here enabled a detailed study, under ambient conditions, of the magnetostatic response and thermally-assisted driven vortex core hopping dynamics in a mesoscopic permalloy element under applied field. This torque magnetometry technique complements other device based nanoscale magnetic probes. Compared to planar micro-Hall approaches \cite{ref:rahm2003vpi}, which have been used to probe single pinning sites but have not been used to measure RF susceptibility, nanocavity torque magnetometry offers higher frequency operation. While it has yet to offer the single spin sensitivity of NV centre based imaging \cite{ref:tetienne2014nic, ref:rugar2015pmr}, it provides comparatively fast acquisition of net magnetization, allowing measurement of magnetic hysteresis and susceptibility. Reconfiguration of the RF fields allows probing of enhanced susceptibility components of single  pinning events, and demonstrates that this magnetometry approach fulfills key requirements for an optomechanical lab-on-a-chip for nanomagnetism.

%%%%%%%%%%%%%
\newpage

%%%% BIBLIOGRAPHY %%%%
\bibliographystyle{nature}
\bibliography{nano_bib}

\section*{Acknowledgments}
This work is supported by the Natural Science and Engineering Research Council of Canada (NSERC), Canada Research Chairs, the Canada Foundation for Innovation (CFI), and Alberta Innovates Technology Futures (AITF).  Many thanks to Aaron Hryciw, Matthew Mitchell, Miroslav Belov, and David Fortin for their technical contributions. We also thank the staff of the nanofabrication facilities at the University of Alberta and at the National Institute for Nanotechnology as well as the machinists at the University of Calgary Science Workshop for their technical support.

%%% AUTHOR CONTRIBUTIONS %%%
\section*{Author contributions}
P.\ E.\ B.\ and M.\ R.\ F.\ conceived and supervised the project. M.\ W.\ and N.\ L.\ W.\ designed and fabricated the devices. N.\ L.\ W.\ imaged the devices. M.\ W.\ set up the measurement equipment including the fiber taper. M.\ W., N.\ L.\ W.\ and T.\ F.\ performed measurements on the device. M.\ W., N.\ L.\ W., T.\ F.\ and F.\ F.\ analyzed the data. M.\ W.\ and N.\ L.\ W.\ prepared figures. F.\ F.\ and T.\ F.\ contributed simulations to the manuscript. F.\ F.\ helped with the theoretical framework for the RF susceptibility mixing scheme in the supplementary material. J.\ E.\ L.\ provided guidance and technical assistance with instrumentation and measurements. All co-authors contributed to and proofread the manuscript.

%%% ADDITIONAL INFORMATION %%%
\section*{Additional information}
Supplementary information is available in the online version of the paper. Reprints and permission information is available online at \url{www.nature.com/reprints}. Correspondence and requests for materials should be addressed to P.\ E.\ B.\ (pbarclay@ucalgary.ca) and M.\ R.\ F. (freemanm@ualberta.ca).

%%% COMPETING FINANCIAL INTERESTS %%%
\section*{Competing financial interests}
The authors declare no competing financial interests.

%%% METHODS %%%
\section*{Methods}

\textbf{Permalloy deposition.} Permalloy structures with a thickness of 40 nm were patterned onto undercut SBC devices using ultra-high vacuum collimated deposition and a lift-off process \cite{ref:diao2013sfl}. The pad of area 1.4 $\times$ 1.3 $ \mu \text{m}^2$ is partially covered with permalloy due to imperfect lithographic alignment during the lift-off process, resulting in the ``mushroom'' shape of the island.  Because polycrystalline permalloy is optically absorbing, the permalloy island is positioned far from the nanocavity center, where it does not degrade $Q_\text{o}$ by interacting directly with the nanocavity optical mode.

\textbf{Measurement setup.} To perform nanocavity-optomechanical torque magnetometry, a permanent magnet (N50 neodymium iron boron, 2.5 cm$^3$) was mounted on a motorized stepper rail and used to apply a stable and finely adjustable $H^\text{DC}_\text{x}$. The field magnitudes were recorded with a three-axis Hall probe placed below the sample chip (Sentron 3M12-2). The RF coil positioned beneath the sample chip was used for generating $H^\text{RF}_\text{z}$ and $H^\text{RF}_\text{x}$. This coil was integrated into an optical fiber taper probing setup identical to that used in previous nanophotonic cavity optomechanics experiments \cite{ref:wu2014ddo}. The schematic of the setup presented in Fig.\ \ref{fig:Sensitivity}a illustrates the detection of the nanobeam motion through a dimpled optical fiber taper (more details in \href{\SMlink}{Supplement} \S1). The dimple is positioned in contact with the top surface of the fixed nanobeam such that in the vicinity of the nanocavity gap region, the fiber taper is aligned $<$ 200 nm from the device, where it induces significant dispersive optomechanical coupling \cite{ref:hryciw2015tno} and evanescently couples light into an out of the nanocavity.  The optical transmission of a tunable laser (Santec TSL-510, wavelength range 1500 nm to 1630 nm) source through the fiber taper was detected using a low-noise photodetector (New Focus 1811) and analyzed using a real-time spectrum analyzer (Tektronix RSA 5103B) and lock-in amplifier (Zurich Instruments HF2LI).  For driving the RF coil, a reference tone was passed from the lock-in amplifier through an RF power amplifier (ENI 403L, 37 dB gain).   All measurements were conducted at ambient temperature and pressure within a nitrogen-purged environment. 

\textbf{Micromagnetic simulations.} Landau-Lifshitz-Gilbert-based micromagnetic simulations were performed with MuMax 3.5 GPU-accelerated open-source software \cite{ref:vansteenkiste2014dvm}, using a three dimensional grid size of 5 nm and characteristic thin film properties of permalloy: saturation magnetization $M_\text{S} $ = 780 kA/m and an exchange stiffness constant $A_\text{ex}$ = 13 pJ/m.  From the calibration of a similar permalloy film (under the same conditions), an experimental value of $M_\text{S} $ = 770 kA/m was obtained. Since the calibrated film was not deposited at the same time as the permalloy pad under study here however, there is some uncertainty in the value of $M_\text{S}$.  The Gilbert damping constant was set to $\alpha$ = 1 to minimize the simulation time required for the quasi-static hysteresis. From the simulations, the net magnetization of the structure at an applied field of 45 kA/m was found to be $M = 0.965\thinspace M_\text{S}$, and this number was assumed also to be representative for the experiment. 

\end{document}